\documentclass[prd,aps,preprintnumbers,amssymb]{revtex4}
\usepackage{graphicx}
\usepackage{dcolumn}
\usepackage{bm}
\begin{document}
\title{
QED\ Near the Decoupling Temperature}
\author{Samina S. Masood} 
\email{masood@uhcl.edu}
 \affiliation{Department of Physics, University of
Houston Clear Lake, Houston TX 77058}
\begin{abstract}
We study the effective parameters of QED near the decoupling temperature and show that the QED perturbation theory works perfectly fine at temperatures,
below the decoupling temperature. Temperature dependent selfmass of electron, at $T=m$, gives two different values when approached to the same overlapping point.  It ia shown that at $T=m$, change in thermal contribution of the electron selfmass is 1/3 of the low temperature value and 1/2 of the high temperature value. 
The difference of behavior measures the electron background contributions at $T=m$.
These electrons are emitted through beta decay. 
This rise in mass affects the QED parameters and change the electromagnetic properties of the medium with temperature also. However, these contributions are ignorable near the decoupling temperature.  
\end{abstract}
  
\maketitle


\section{Introduction}

\bigskip 

Renormalization techniques of perturbation theory are used to calculate
temperature dependence of renormalization constants of QED (quantum
electrodynamics) at finite temperature [1-19]. The values of electron mass,
charge and wavefunction, at a given temperature, represent the effective
parameters of QED at those temperatures. The magnetic moment of electron
[9,18], dynamically generated mass of photon [4,9,15,19] and QED coupling
constants are estimated as functions of temperature. Moreover, thermal
contributions to the electric permittivity, magnetic permeability and
dielectric constant of a medium can be obtained from the photon selfmass.
Some of the important parameters of QED plasma such as Debye shielding
length, plasma frequency and phase transitions can also be determined from the
properties of the medium itself.

In this paper, we quantitatively analyze the existing results of temperature
dependent renomalization constants. The renormalization scheme of QED in
real-time formalism, is used to calculate the electron mass, wavefunction and
charge of electron. It is now well-known that the existing first order thermal 
corrections to the renormalization constants give the quadratic 
dependence of QED parameters on temperature T, expressed in units of electron mass m. 
The existing scheme of calculations is very useful and it works perfectly fine below the
decoupling temperature,i.e.; 2MeV. It can be explicitly checked that in
the existing scheme of calculation, QED theory remains renormalizeable at $%
T\lesssim 5MeV$. However, the first order corrections will exceed the original
values of QED parameters at much higher temperatures and hard thermal loops
will not appear before that. Then we have to look for a new scheme of calculations for
temperatures much higher than the decoupling temperatures. There are already
developed methods [20,21], which could be used for this purpose. However,
below the neutrino decoupling temperature, the real part of the propagators
serve the purpose.

   Existing analytical results are based on the methods of perturbation
theory in vacuum and are extremely useful to explain the behavior of QED up
to the decoupling temperature. Renormalization methods of QED at finite
temperature ensures a divergence free QED in a thermal medium below the
decoupling temperature. Without going in to the calculational details, we
give a brief overview of the existing results in the real-time formalism. It
is possible to separate out the temperature dependent contributions from the
vacuum contribution as the statistical distribution functions contribute
additional statistical terms, both to fermion and boson propagators, in the
form of Fermi-Dirac distribution and Bose-Einstein distribution functions,
respectively. The Feynman rules of vacuum theory are used with the
statistically corrected propagators given as

\begin{equation}
D_{\beta }(k)=\frac{i}{k^{2}-m^{2}+i\varepsilon }+\frac{2\pi }{e^{\beta
E_{k}}-1}\delta (k^{2}-m^{2}),
\end{equation}%
for bosons, and \bigskip

\begin{equation}
S_{F}(p)=\frac{i}{\not{p}-m+i\varepsilon }-\frac{2\pi(\not{p}+m)}{e^{-\beta
E_{p}}+1}\delta (p^{2}-m^{2}),
\end{equation}%
for fermions.

The electron mass, wavefunction and charge are then calculated in a
statistical medium using Feynman rules of QED, with the modified propagators
given in Eqs.(1) and (2). Since the temperature corrections are additive
corrections in the propagator and appear as additive terms in the matrix element,
 thermal radiative corrections can be studied independent of vacuum
corrections at the one loop level. We restrict ourselves to the one loop
contributions only as it can be easily shown that the higher order
contributions [13-19] are smaller than the first order contributions at
these temperatures and ensure the validity of the renormalization scheme.

\section{Selfmass of Electron}

The renormalized mass of electrons $m_{R}$\ can be represented as a physical
mass $m_{phys}$of electron and is defined in a hot and dense medium as,%
\begin{equation}
m_{R}\equiv m_{phys}=m+\delta m(T=0)+\delta m(T).
\end{equation}%
where $m$ is the electron mass at zero temperature and $\delta m(T=0)$\
represents the radiative corrections from vacuum and $\delta m(T)$ are the
contributions from the statistical background at nonzero temperature T. The
physical mass can get radiative corrections at different orders of $\alpha $
and can be written as:%
\begin{equation}
m_{phys}\cong m+\delta m^{(1)}+\delta m^{(2)}+....
\end{equation}%
where $\delta m^{(1)}$ and $\delta m^{(2)}$\ are the shifts in the electron
mass in the first and second order in $\alpha $, respectively. The physical
mass is deduced by locating the pole of the fermion propagator $\frac{i(\not%
{p}+m)}{p^{2}-m^{2}+i\varepsilon }$ in thermal background. For this
purpose, we sum over all the same order diagrams. Renormalization is
established by demonstrating the order-by-order cancellation of
singularities. All the finite terms from the same order in $\alpha $ are
combined together to evaluate the same order contribution to the physical
mass given in eq.(4). The physical mass in thermal background, up to order $%
\alpha ^{2}$ [13-19], is calculated using the renormalization techniques of
QED. Writing the selfmass of electron [1,6] as:%
\begin{equation}
\Sigma (p)=A(p)E\gamma _{_{0}}-B(p)\vec{p}.\vec{\gamma}-C(p),
\end{equation}%
where $A(p)$, $B(p)$, and $C(p)$ are the relevant coefficients that are
functions of electron momentum only. Taking the inverse of the propagator
with momentum and mass terms are separated as:%
\begin{equation}
S^{-1}(p)=(1-A)E\gamma ^{o}-(1-B)p.\gamma -(m-C).
\end{equation}%
The temperature-dependent radiative corrections to the electron mass up to
the first order in $\alpha $, are obtained from the temperature dependent
propagators as%
\begin{equation}
m_{phys}^{2}\equiv E^{2}-|\mathbf{p}|^{2}=m^{2}\left[ 1-\frac{6\alpha }{\pi }%
b(m\beta )\right] +\frac{4\alpha }{\pi }\left[ mT\text{ }a(m\beta )+\frac{2}{%
3}\alpha \pi T^{2}-\frac{6}{\pi ^{2}}c(m\beta )\right].
\end{equation}%
giving

\[
\frac{\delta m}{m}\simeq \frac{1}{2m^{2}}\left( m_{phys}^{2}-m^{2}\right)\\ 
\]
\begin{eqnarray}
 \frac{\delta m}{m} \simeq  \frac{\alpha \pi T^{2}}{3m^{2}}\left[ 1-\frac{6}{\pi ^{2}}c(m\beta )%
\right] +\frac{2\alpha }{\pi }\frac{T}{m}a(m\beta )-\frac{3\alpha }{\pi }%
b(m\beta ).
\end{eqnarray}%

where $\frac{\delta m}{m}$ is the relative shift in electron mass due to
finite temperature which was determined in Ref. [7] with%

\begin{equation}
a(m\beta )=\ln (1+e^{-m\beta }),
\end{equation}

\begin{equation}
b(m\beta )=\sum_{n=1}^{\infty }(-1)^{n}Ei(-nm\beta ),
\end{equation}%

\begin{equation}
c(m\beta )=\sum_{n=1}^{\infty }(-1)^{n}\frac{e^{-nm\beta }}{n^{2}}.
\end{equation}%
 
The validity of Eq.(4) can be ensured for $T \leq 2MeV$, the decoupling temperature, as $\frac{\delta m}{m}$
is always smaller than unity within this limit. This scheme of calculations
will not work for higher temperatures, as the summation over all orders of perturbative correction may exceed the original values of QED parameters, after $5MeV$. We will
have to develop a new method of calculations for the higher
temperature limit. At low temperature $T<m$, the functions $a(m\beta )$, $%
b(m\beta )$, and $c(m\beta )$ fall off in powers of $e^{-m\beta }$
in comparison with $\left( \frac{T}{m}\right) ^{2}$ and can be neglected in
the low temperature limit giving,%

\begin{equation}
\frac{\delta m}{m}(T< m)=\frac{\alpha \pi T^{2}}{%
3m^{2}}.
\end{equation}%
In the high-temperature limit, $a(m\beta )$ and $b(m\beta )$ are vanishingly
small whereas $c(m\beta )\longrightarrow -\pi ^{2}/12$, yielding%
\begin{equation}
\frac{\delta m}{m}({T> m})=\frac{\alpha \pi T^{2}}{2m^{2}}.
\end{equation}%
Eqs.(12) and (13) give $\frac{\delta m}{m}=7.647\times 10^{-3}\frac{T^{2}}{%
m^{2}}$ for low temperature and $\frac{\delta m}{m}=1.147\times 10^{-2}\frac{%
T^{2}}{m^{2}}$ for high temperature showing that the rate of change of $%
\delta m$ is larger at $T>m$ as compared to $T<m$. Subtracting eq.(12) from
(13), the change in $\frac{\delta m}{m}$ between low and high temperature
ranges can be written as

\begin{equation}
\Delta(\frac{\delta m}{m})=\pm \frac{\alpha \pi T^{2}}{6m^{2}}=\pm
3.8\times 10^{-3}\frac{T^{2}}{m^{2}}
\end{equation}%
showing that the $\Delta (\frac{\delta m}{m})=\pm 3.8\times 10^{-3}$at $T=m$%
.\ 
It can be easily checked that the low temperature behavior will give a
 50\% decrease in selfmass as compared to high T value.
Whereas the high T behavior will give 33\% more selfmass as compared to the
low T value. This difference is due to the fact that at low temperature, only hot boson contribution is calculated using Eq.(12), whereas Eq.(13) includes the fermion background contribution also. Eq.(14) is a measure of fermion background contribution that is ignoreable at $T<m$ but cannot be ignored at $T>m$. This difference keeps on increasing with temperature also. 

Temperature dependence of QED parameters is a little more
complicated and significant because of the change in
matter composition, during nucleosynthesis. Therefore Eqs.(7) and (8) are
required for $T\sim m$ region and help to compute the change in
thermal behavior of QED parameters due to the change in matter composition, 
carried out by beta decay and other processes, at that time. This difference can be clearly seen in Figure(1). We plot of eqs. (12) and (13), corresponding to low temperature and high temperature and show that both plots start to give a disconnected region near $T\sim m$, i.e; the nucleosynthesis temperature.
Slope of both graphs is also different, indicating that $T\sim m$ induces a
change in thermal properties for a heating and a cooling system.

\begin{figure}[!hth]
\begin{center}
  \begin{tabular}{c}
    \mbox{\includegraphics[width=3.5in,angle=0]
{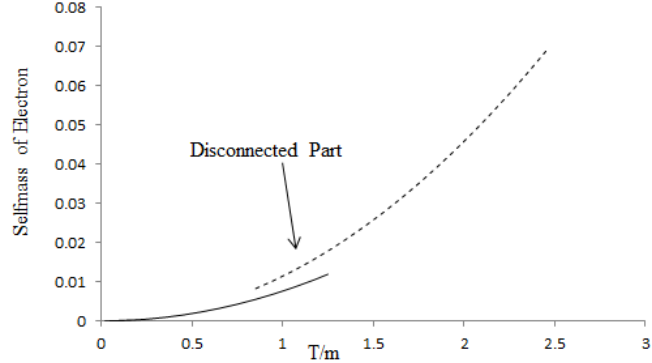}}\\
\end{tabular}
\end{center}
\caption{ 
Low temperature values of selfmass of electron (solid line) and high temperature values (broken line) are plotted as a function of temperature T. Around T=m, both of the graphs approach to different values. }
\label{fig1}
\end{figure}
The disconnected region in Figure (1) shows that the $\frac{T^{2}}{m^{2}}$
behavior is needed to be modified to find the missing link between the low
temperature and high temperature behavior. Eq.(8) provides the information
about the disonnected region and eqs. (12) and (13) can be derived from
eq.(8). Figure (1) also shows that the thermal corrections to the electron mass are not
significant below $T\sim $ $0.05 MeV$.

The modification in the electron mass behavior in the range $T\sim m$, is
estimated by Eq. (12). It is also clear from Eq. (13) that after $5MeV$%
, temperature dependence correction term can easily reach ($\frac{\delta m}{m}>1$), even at
the one-loop level. Higher order corrections [13-16] will grow up rapidly
at high temperatures as we can approximate it as: 

\begin{equation}
m_{phys}=m\left[ 1+\frac{\delta m}{m}+\frac{1}{2}(\frac{\delta m}{m}%
)^{2}+....\right] \approx m\exp (\frac{\delta m}{m})
\end{equation}

for the temperature T in the units of the elctron mass m as $0.1m\leq T\geq 10m$. This almost exponential growth in the physical mass is very important here.

Figure (1) shows that a change in the QED behavior occurs around $T \sim m$ and is clearly related to nucleosynthesis. Right after decoupling, beta decay and other processes involving the electron mass, change the composition of matter and electron picks up thermal mass from hot fermion
loop. At $T>5$ MeV, the renormalization scheme of perturbative QED may not
be valid as beta decay contribute through weak interactions. 

\section{\textbf{Wavefunction Renormalization:}}

The electron wavefunction in QED\ is related to the selfmass of electron
through Ward identity. The factor ($1-A$) is required for renormalization,
because then the propagator can also be renormalized by replacing

\[
\frac{1}{\not{p}-m+i\varepsilon }\rightarrow \frac{Z_{2}^{-1}}{\not
{p}-m+i\varepsilon }.
\]

Thus, for Lorentz invariant self-energy, the wavefunction renormalization
constant can equivalently be expressed as
\[
Z_{2}^{-1} = 1-A
\] 
\begin{eqnarray}
Z_{2}^{-1} =1-\frac{\partial \Sigma (p)}{\partial \not{p}}.
\end{eqnarray}
The fermion wavefunction renormalization in the finite temperature field
theory can be obtained in a similar way as discussed in vacuum theories.
However, the Lorentz invariance in the finite temperature theory is imposed
by setting $A=B$ in Eq.(5). Thus, using Eqs. (16) and (5), one obtains [7]

\[
Z_{2}^{-1}(m\beta ) =Z_{2}^{-1}(T=0)-\frac{2\alpha }{\pi }\int_{0}^{\infty
}\frac{dk}{k}n_{_{B}}(k)-\frac{3\alpha }{\pi }b(m\beta ) 
\]
\begin{eqnarray}
 +\frac{\alpha T^{2}}{\pi vE^{2}}\ln \frac{1+v}{1-v}{ \{}\frac{\pi
^{2}}{6}+m\beta a(m\beta )-c(m\beta ){ \}}.
\end{eqnarray}

giving the low temperature values as

\begin{equation}
Z_{2}^{-1}=Z_{2}^{-1}(T=0)-\frac{2\alpha }{\pi }\int \frac{dk}{k}n_{B}(k)+%
\frac{\alpha \pi T^{2}}{6E^{2}}\frac{1}{v}\ln \frac{1-v}{1+v}
\end{equation}

and high temperature value as

\begin{equation}
Z_{2}^{-1}=Z_{2}^{-1}(T=0)-\frac{2\alpha }{\pi }\int \frac{dk}{k}n_{B}(k)+%
\frac{\alpha \pi T^{2}}{4E^{2}}\frac{1}{v}\ln \frac{1-v}{1+v}
\end{equation}

For small values of $v$, the low and high temperature values can be
determined from eqs. (18) and (19) as 

\begin{equation}
Z_{2}^{-1}=Z_{2}^{-1}(T=0)-\frac{2\alpha }{\pi }\int \frac{dk}{k}n_{B}(k)-%
\frac{\alpha \pi T^{2}}{3E^{2}}
\end{equation}

for low temperature, and

\begin{equation}
Z_{2}^{-1}=Z_{2}^{-1}(T=0)-\frac{2\alpha }{\pi }\int \frac{dk}{k}n_{B}(k)-%
\frac{\alpha \pi T^{2}}{2E^{2}}
\end{equation}

for high temperature.

The finite part of eqs. (20) and (21) is equal to $\frac{\delta m}{m}$ at the lowest value of energy, i.e.;
 $E=m$ in the relevant temperature range. These terms are suppressed at large
value of electron energy E as they are suppressed by a factor $\frac{T^{2}}{
E^{2}}$. However, the calculated value at that temperature is significantly
different. The difference in the thermal contribution can easily be found to
be about 50\% of the low temperature value and around 33.3\% of the high
temperature value. This difference can be mentioned as

\begin{equation}
\Delta (Z_{2}^{-1})\approx \frac{\alpha \pi T^{2}}{6E^{2}}=3.8\times 10^{-3}%
\frac{T^{2}}{E^{2}}
\end{equation}

and is similar to the selfmass corrections. However, a difference of sign is noticeable, 
showing the decrease in the renormalization constant and not the increase. However, 
the finite part of the wavefunction renormalization constant can be obtained
by finding a ratio of temperature with the Lorentz energy E. The minimum
value of this energy is equal to mass. Following eq.(15), the higher order
contributions to the wavefunction can then be approximated as

\begin{equation}
Z_{2}^{-1}=Z_{2}^{-1}(T=0)-\frac{2\alpha }{\pi }\int \frac{dk}{k}%
n_{B}(k)-\exp (\frac{\alpha \pi T^{2}}{2E^{2}})
\end{equation}
 \begin{figure}[!hth]
\begin{center}
  \begin{tabular}{c}
    \mbox{\includegraphics[width=3.0in,angle=0]
{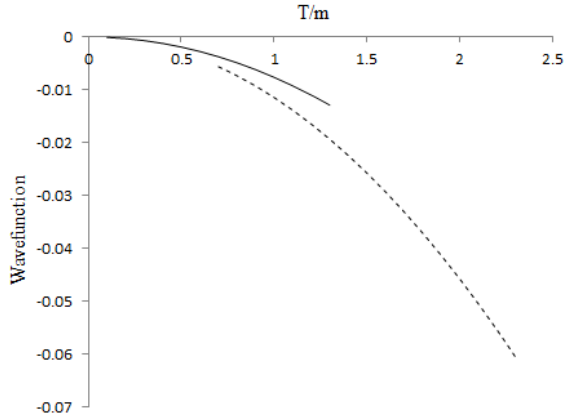}}\\
  \end{tabular}
\end{center}
\caption{
Low temperature value (broken line) and high temperature value (solid line) of the electron wavefunction renormalization constant are plotted as a function of T. 
}
\label{FIG2}
\end{figure}
As shown in Figure 2, the wavefunction renormalization contributions are negative. 
It shows that the low temperature ($T<m$) contribution is simply ignorable as compared to 
high temperature ($T>m$) contribution everywhere below the decoupling temperature.  
The finite term can be ignorable at $T=m$, and even at the temperatures
higher than nucleosynthesis as $T<E$ at those temperatures because E is
always greater than m. For large E, thermal contributions are even smaller and more
ignorable. So the two interesting physical limits give
smaller thermal contribution in electron wavefunction as the relevant
temperature limits can be defined as $m<T<E$ and $T<m<E$, which ensures the
renormalizability of QED at comparatively higher temperature as compared to
selfmass. 

We do not discuss the infrared singularity term as it has already been
studied in literature. We are interested to compare the finite contributions
only to see how they change in different regions of temperatures can justify the validity of renormaliztion scheme of vacuum theories. 

\section{Photon Selfmass and QED Coupling Constant}

Selfmass of photon and the electron charge also behave differently for a
cooling and a heating system at high temperature. The electron charge is not expected to be changed due to the presence of neutral photons. That is the reason that the
electron charge and the coupling constant does not show significant
temperature dependence for $T<m$. However, they have significant thermal
contributions at high temperatures ($T>m$). Difference of behavior of a 
cooling and a heating system starts to be significant near $T\sim m$ from the low temperature side 
and near decoupling temperature from the high temperature side. This difference in the
coupling constant occurs due to the dynamically generated mass of photon that couples with the hot electrons in thermal medium. The emission of electrons looks more natural due to the beta decay processes during nucleosynthesis.

Calculations of the vacuum polarization tensor [7] show that the real part of
the longitudinal and transverse polarization components of the
polarization tensors can be evaluated, in the limit $\omega \rightarrow 0$,
as:
\begin{equation}
Re\Pi _{L}^{\beta }\left( k,0\right) = \frac{4 {\pi}{\alpha} }{3}%
\left \{ T^{2}+\frac{k^{2}}{2\pi ^{2}}\ln \frac{m}{T}  \right \} , \\
Re\Pi _{T}^{\beta }\left( k,0\right) = \frac{2\alpha }{3\pi }%
\left \{ k^{2}\ln \frac{m}{T}+.....\right \}.\qquad \text{\ }
\end{equation}
giving the interaction potential, in the rest frame of the charged particles
as
\begin{equation}
V(k)\equiv e_{R}^{2}\delta _{\mu 0}\left[ \frac{u_{\mu }{}u_{\nu }}{k^{2}-%
\frac{4\pi \alpha }{3}\left \{ T^{2}+\frac{k^{2}}{2\pi ^{2}}\ln \frac{m}{T}%
\right \} }+\frac{g_{\mu \nu }-u_{\mu }{}u_{\nu }}{k^{2}-\frac{2\alpha }{%
3\pi }k^{2}\ln \frac{m}{T}}\right],
\end{equation}
\begin{equation}
V(k)=e_{R}^{2}\delta _{\mu 0}\Delta _{\mu \nu }\delta _{\nu 0},
\end{equation}

$e_{R}$\ is the renormalized charge at $T=0$. $V(k)$ can be expanded, at low
temperature as:

\begin{equation}
V(k)\equiv e_{R}^{2}\left( 1+\frac{2\alpha }{3\pi }\ln \frac{T}{m}\right) %
\left[ \frac{u_{0}^{2}}{k^{2}+\frac{4\pi \alpha T^{2}}{3}}+\frac{%
1-u_{0}^{2}{}}{k^{2}}\right] .
\end{equation}

The constant in the longitudinal propagator is the plasma screening mass,
therefore, whole of the outside factor corresponds to the charge
renormalization and in turn to the coupling constant. We may then write the
coupling constant at low temperature as [1-3]:

\begin{equation}
\alpha \left( T\right) =\alpha \left( T=0\right) \left( 1+\frac{2\alpha }{%
3\pi }\ln \frac{T}{m}\right) .=\alpha \left( T=0\right) \left( 1+1.55\times
10^{-3}\ln \frac{T}{m}\right) 
\end{equation}%
The factor $1.55\times 10^{-3}\ln \frac{T}{m}$ is a slowly varying function
of temperature and does not give any significant contribution near the
decoupling temperature and remains insignificant for a large range of
temperature.

The temperature dependent factor in the longitudinal propagator $\frac{4\pi
\alpha T^{2}}{3}$ is the plasma screening frequencies or selfmass of photon
that contribute to the QED coupling constant at finite temperature.

\bigskip For generalized temperatures, the charge renormalization constant $%
Z_{3}$ can be written as [9]
\begin{equation}
Z_{3}=1-\frac{2e^{2}}{\pi ^{2}}{ \{}\frac{c(m\beta )}{\beta ^{2}}-%
\frac{ma(m\beta )}{\beta }\ -\frac{1}{4}{ (}m^{2}-\frac{\omega ^{2}}{3}%
{)}b(m\beta )\}.
\end{equation}

Also, the electric permittivity is
\begin{equation}
{\varepsilon}(K)\simeq 1+\frac{4e^{2}}{\pi ^{2}K^{2}}\left( 1-%
\frac{\omega ^{2}}{k^{2}}\right) {}\left( 1-\frac{\omega }{2k}\ln 
\frac{\omega+k}{\omega -k}\right) \left( \frac{c(m\beta )}{\beta ^{2}}-%
\frac{ma(m\beta)}{\beta }\right) \quad
\\
-\frac{1}{4}\left(2m^{2}-\omega ^{2}
+\frac{11k^{2}+37\omega^{2}}{72}\right)b(m\beta){\}},
\end{equation} 

and the magnetic permeability is%
\[
\frac{1}{\mu (K)} \simeq 1+\frac{2e^{2}}{\pi ^{2}k^{2}K^{2}}{[}%
\omega ^{2}\left \{ 1-\frac{\omega ^{2}}{k^{2}}-\left( 1+\frac{k^{2}}{\omega
^{2}}\right) \left( 1-\frac{\omega ^{2}}{k^{2}}\right) \frac{\omega }{2k}\ln 
\frac{\omega+k}{\omega -k}\right \} 
\]
\begin{eqnarray}
\times \left( \frac{c(m\beta )}{\beta ^{2}}-\frac{ma(m\beta )}{\beta }%
\right) -\frac{1}{8}\left( 6m^{2}-\omega ^{2}+\frac{129\omega ^{2}-109k^{2}}{%
72}\right) b(m\beta )].
\end{eqnarray}
In the limit $T>m$, the wavefunction renormalization constant can be written
as:

\begin{equation}
Z_{3}=1+\frac{\alpha T^{2}}{6m^{2}}
\end{equation}

giving the renormalized coupling constant as

\begin{equation}
\alpha =\frac{e^{2}/(\not{h}c)}{4\pi \epsilon _{0}}(1+\frac{\alpha T^{2}}{%
6m^{2}})=\frac{\mu _{0}e^{2}c}{2h}(1+\frac{\alpha T^{2}}{6m^{2}})
\end{equation}

Eq. (32) gives $Z_{3}=1+1.2\times 10^{-3}\frac{T^{2}}{m^{2}}$ and leaves the
perturbation series valid for at least $T\leq 10m.$

It is clear from eqs. (29-32) that the coupling constant is basically
changed from the hot fermion loop contributions. Hot bosons do not change
the coupling constant and the vacuum fluctuations occur due to fermion loops
at the first order in $\alpha $.

Temperature corrections to the coupling constant start to become noticeable
during nucleosynthesis. However, the value of $c(m\beta )$ approaches a
constant value $\frac{-\pi ^{2}}{12}$ at $T\sim m$ and start to become
significant at higher temperatures. After $T>5MeV$, thermal contributions
indicate that the coupling constant can grow larger than unity at high
temperature indicating a problem for perturbative behavior of QED. We need
to use non-perturbative methods to establish renormalization of QED at those
temperatures.
\bigskip
\section{Results and Discussion}

\bigskip

\begin{table}
\begin{center}
  \begin{tabular}{c}
   \mbox{\includegraphics[width=3.0in,angle=0]
{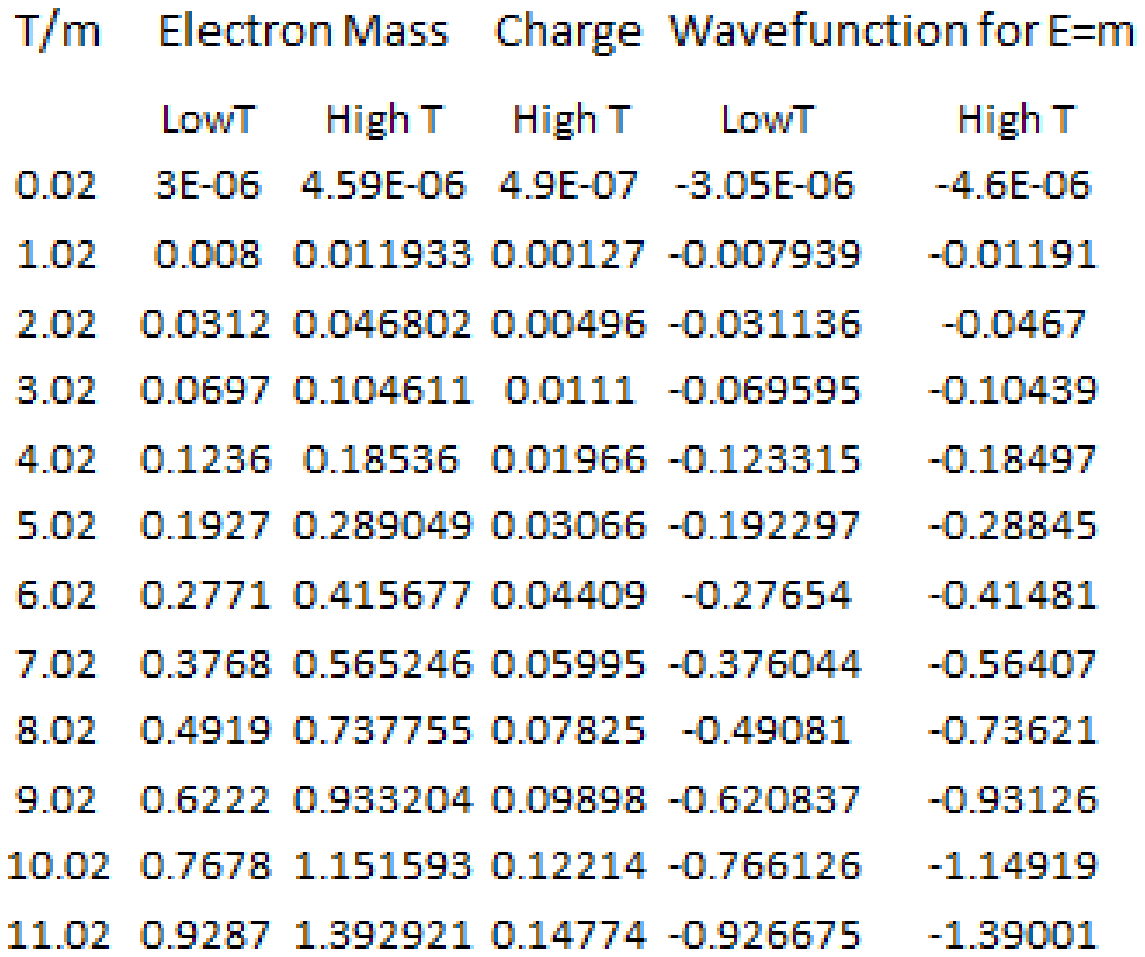}}\\
  \end{tabular}
\end{center}
\caption{Renormalization constants of QED at finite temperature corresponding to the $T<m$ and $T>m$ expressions.
} \label{Table1}
\end{table}


Quantitative study of renormalization constants at finite temperature [Table 1] 
shows that all the renormalization constants are definitely finite below 5MeV (i.e.;$T<10m$). 
Table 1 indicates that the first order corrections to all the renormalization constants of QED are not always ignorable, and significantly grows around the decoupling temperature. 
Computation of thermal contributions of the renormalization constants, around the decoupling temperature, 
is not straightforward. In this range, the largest thermal contribution comes from the electron selfmass.
However, thermal contribution to the wavefunction renormalization constant is small at these temperatures,
 and high energies ($E>>m$). Eqs. (18) and (19) show that the maximum thermal contribution to electron wavefunction renormalization constant is equal to the selfmass of electron at $E= m$ and that is the largest contribution.
There is no low temperature contributions to electron charge, as well as the QED\
coupling constant, due to the absence of hot fermions in the medium. 
Hot fermion loop contributions are ignorable at low temperatures $(T<m)$. However, at $T>m$ the coupling constant starts to pick up thermal corrections, due to the dynamically generated mass of photon, which is generated through the fermion background only.

 \begin{figure}[!hth]
\begin{center}
  \begin{tabular}{c}
    \mbox{\includegraphics[width=4.5in,angle=0]
{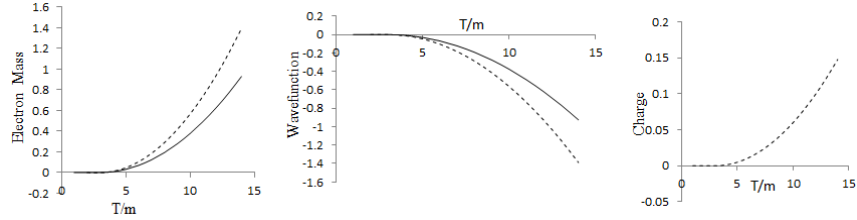}}\\
  \end{tabular}
\end{center}
\caption{
Low temperature values of the renormalization constants of QED (solid line) and high temperature values (broken line) are plotted as a function of temperature T. Around low temperatures, both of the graphs approach to different values.
} 
\label{FIG3}
\end{figure}

A comparison of the statistical background contributions to different renormalization constants shows that, in this scheme of calculations, the selfmass corrections are the largest corrections at all temperatures. Wavefunction renormalization is significant at low energies but the coupling constant does not seem to get significant thermal corrections at decoupling temperature.
This scheme of calculation is very helpful to compute QED parameters during nucleosynthesis in terms of $a(m\beta )$, $b(m\beta )$
and $c(m\beta )$. Figure 3 gives a plot of thermal contributions to electron
mass ($\frac{\delta m}{m}$ ), wavefunction renormalization constant
($Z_{2}^{-1}+\frac{2\alpha }{\pi }\int \frac{dk}{k}n_{B}(k)$ ) at low energy ($E=\sqrt{2}m$) and electron charge ($Z_{3}-1$ ), that can be derived for $T<<m$ and $T>>m$
ranges from the same equation. Since we are dealing with the exponential
functions in this study, an order of magnitude difference is a safe limit
for these approximations. It means $T\leq 0.1 m$ is $T<< m$ and $T\geq 10 m$ is
 $T>>m$. The broken line is high temperature limit ($T\geqslant m$ )
and the solid lines correspond to the low temperature ($T
\leq $ $m$ ) limit of the corresponding parameters. Both limits are plotted 
for overlapping temperatures to show that the difference
between the low temperature and the high temperature values is due to
the contributions of hot fermion loops at high temperatures. Contributions
of function $c(m\beta )$ vanishes at low T and it sums up to (-$\frac{%
\pi ^{2}}{12}$) for large T values. This function actually bring the fermion
loop contributions at high T when more fermions are generated during the
nucleosynthesis and their presence is not ignorable in the medium. Separations between two curves in the above graphs of Figure 3, measure the fermion contributions at high temperatures. However, around the decoupling temperature, this separation indicate complicated processes which cause the emission of electrons in beta decay and even absorption, during nucleosynthesis. These electrons acquire thermal equilibrium with the medium. However, $a(m\beta )$, $b(m\beta )$ and $c(m\beta )$ functions are needed to study the behavior of these parameters near $T\sim m $ and explain the disconnected region of Figure (1). 

Plot of these renormalization constants, show that the temperature
corrections are to small to differentiate between the boson and fermion contributions, for T sufficiently smaller than m. Significant thermal behavior starts near $T\sim m$, as indicated in Figures 3.

Cooling universe of the standard big bang model behaves differently
after the neutrino decoupling. Nucleosynthesis starts right after the
neutrino decoupling and the helium synthesis takes place when the
temperature of the universe is cooled down to the temperature of electron
mass. This is actually a period, where the finite temperature corrections to
QED\ parameters are significant and complicated enough to evaluate it
numerically. However, the temperature dependent QED parameters are helpful to
describe the observations of WMAP (Wilkinson Microwave Anisotropy Probe) data [22-23]. 
After the nucleosynthesis is complete, all the renormalization constants depend quadratically on temperature, though it may not always be significant.
 At high temperatures, QED coupling plays its role in modifying QED parameters for nucleosynthesis. With the help of
these effective parameters of QED, the abundance of helium in the early
universe can be estimated [12] precisely at a given temperature. The
temperature dependent QED corrections to the nucleosynthesis parameters
improve the results of standard big bang model of cosmology and testing of
the standard model with WMAP becomes more reasonable. The same techniques can even be used to calculate
perturbative effects in QCD [24] and electroweak processes [25] at low
temperatures.
\bigskip \bigskip 

\textbf{REFERENCES }

\begin{enumerate}
\item See for example: Samina Masood,`QED at Finite Temperature and
Density.',Lambert Academic Publication, (March,2012)

\item J. F. Donoghue and B. R. Holstein, Rev. \textbf{D28}: 340(1983) ;
[Erratum ibid \textbf{29}:3004 (1983) ].

\item J. F. Donoghue, B. R. Holstein, and R. W. Robinett, Ann. Phys. (N.Y.)
\textbf{164}: 233(1985) .

\item A. E. I. Johansson, G. Peressutti, B. S. Skagerstam, Nucl. Phys.
\textbf{B278}: 324(1986).

\item G. Peressutti, B. S. Skagerstam, Phys. Lett. B110, 406(1982).

\item A.Weldon, Phys. Rev. \textbf{D26}: 1394(1982).

\item K. Ahmed and Samina Saleem (Masood), Phys. Rev. \textbf{D35}:
1861(1987) .

\item K. Ahmed and Samina S. Masood, Phys. Rev. \textbf{D35}, 4020(1987) .

\item K. Ahmed and Samina Saleem (Masood), Ann. Phys. \textbf{207}(N.Y.) 460
(1991) .

\item Samina S. Masood, Phys. Rev.\textbf{\ D44:} 3943(1991).

\item Samina S. Masood, Phys. Rev.\textbf{\ D47}: 648(1993).

\item Samina Saleem ( Masood), Phys. Rev. \textbf{D36}: 2602(1987).

\item Mahnaz Qader, Samina S. Masood, and K. Ahmed, Phys. Rev. \textbf{D44}:
3322(1991) .

\item Mahnaz Qader, Samina S. Masood, and K. Ahmed, Phys. Rev.\textbf{\ D46:}
5633(1992).

\item Samina S. Masood, and Mahnaz Q. Haseeb, Int. J. of Mod. Phys. \textbf{%
A23}, 4709(2008).

\item Mahnaz Q. Haseeb and Samina S. Masood, Chin. Phys. \textbf{C35}: 608
(July 2010).

\item Haseeb M Q and Masood S Samina, Phys. Lett. \textbf{B704:} 66(2011).

\item Samina S. Masood, and Mahnaz Q. Haseeb, `Second Order Corrections to
the Magnetic Moment of Electron at Finite Temperature' arXiv:\textbf{%
1203.3628v2} [hep-th].

\item Samina S. Masood, and Mahnaz Q. Haseeb, `Second Order Photon Loops at Finite Temperature, \textbf{arXiv:1110.3447}[hep-th].

\item Yuko Fueki, Hisao Nakkagawa, Hiroshi Yokota and Koji Yoshida,
Prog.Theor.Phys.\textbf{110}: 777(2003).

\item See for example; N. Fornengo, et. al; Phys. Rev. \textbf{D56}:
5123(1997); Nan Su Commun.Theor.Phys.\textbf{57}:409(2012) and references
therein.

\item E. Komatsu et al. [ WMAP Collaboration], Astrophys. J. Suppl. \textbf{%
192}: 18 (2011); G. Hinshaw et al., Ap. J. Suppl. \textbf{180:} 225 (2009) .

\item Gary Steigman, IAU Symposium No. 265: (2009).

\item Samina S. Masood and Mahnaz Qader Haseeb, Astropart. Phys. \textbf{3}:
405(1995).

\item Samina S. Masood and Mahnaz Qader, Phys. Rev. \textbf{D46}: 511(1992).
\end{enumerate}

\end{document}